\documentclass[letterpaper]{article} 

\usepackage[draft]{aaai25}

\usepackage{times}  
\usepackage{helvet}  
\usepackage{courier}  
\usepackage[hyphens]{url}  
\usepackage{graphicx} 
\usepackage{natbib}  
\usepackage{caption} 
\usepackage{algorithm}
\usepackage{algorithmic}

\usepackage{newfloat}
\usepackage{listings}
\DeclareCaptionStyle{ruled}{labelfont=normalfont,labelsep=colon,strut=off} 
\floatstyle{ruled}
\newfloat{listing}{tb}{lst}{}
\floatname{listing}{Listing}
\title{Lessons for Editors of AI Incidents from the AI Incident Database}
\author {
    Kevin Paeth\textsuperscript{\rm 1},
    Daniel Atherton\textsuperscript{\rm 2,3},
    Nikiforos Pittaras\textsuperscript{\rm 1},
    Heather Frase\textsuperscript{\rm 2,4},
    Sean McGregor\textsuperscript{\rm 1}
}
\affiliations {
    \textsuperscript{\rm 1}UL Research Institutes\\
    \textsuperscript{\rm 2}Responsible AI Collaborative\\
    \textsuperscript{\rm 3}Georgetown University\\
    \textsuperscript{\rm 4}Veraitech\\
    kevin.paeth@ul.org,
    danny@raicollab.org,
    pittarasnikif@gmail.com,
    hnfrase@veraitechus.com,
    sean.mcgregor@ul.org
}

\begin{document}

\maketitle

\begin{abstract}
As artificial intelligence (AI) systems become increasingly deployed across the world, they are also increasingly implicated in \textit{AI incidents} – harm events to individuals and society.
As a result, industry, civil society, and governments worldwide are developing best practices and regulations for monitoring and analyzing  AI incidents.
The AI Incident Database (AIID) is a project that catalogs AI incidents and supports further research by providing a platform to classify incidents for different operational and research-oriented goals.
This study reviews the AIID’s dataset of 750+ AI incidents and two independent taxonomies applied to these incidents to identify common challenges to indexing and analyzing AI incidents.
We find that certain patterns of AI incidents present \textit{structural ambiguities} that challenge incident databasing and explore how \textit{epistemic uncertainty} in AI incident reporting is unavoidable. We therefore report mitigations to make incident processes more robust to uncertainty related to cause, extent of harm, severity, or technical details of implicated systems.
With these findings, we discuss how to develop future AI incident reporting practices.

\end{abstract}

\section{Introduction} \label{introduction}
As practitioners of AI incident reporting over the last seven years, we have acted as subject matter experts to researchers, corporations, and governments seeking to develop domain, company, and country-specific incident databases. Such efforts stand to advance safer AI development and deployment along with impact-centered AI governance. Yet throughout these efforts, there are consistent editorial challenges faced by our database and those databases that have yet to launch.

Incident editing challenges require reflection and accommodation, especially in a time where AI incident reporting requirements are being formalized within legislation. For example, the recent European Union AI Act requires developers and deployers of AI systems implicated in ``serious'' incidents to report to national AI authorities charged with oversight \cite{eu_ai_act}. In the United States, the state of Colorado passed legislation requiring AI system developers and deployers to report instances of algorithmic discrimination \cite{colorado_ai}, and draft legislation is circulating within the AI policy community for national AI incident reporting.

Several corporations are setting up AI-specific incident response programs. These personnel often operate from a history of security incident response. We believe these are positive trends and there is much that can be learned from previously developed incident processes. However, we also believe that the broader community can benefit from a deeper exploration of AI-specific incident cataloging and editing practices.

The contributions of this work are therefore twofold. First, we explore the foundational question of deciding what is and is not an AI incident. While we have contributed to intergovernmental efforts at formalizing the answer, such high level definitions leave many implementation details to the reader when bridging definitions to policy. Second, a collection of incidents is only as useful for preventing incident recurrence as its metadata. Well-motivated analyses of incidents recorded as metadata (i.e., ``taxonomies'' in our parlance) are necessary for making incidents useful for querying and analysis. Within this work, we explore the challenges and solutions for navigating the evolving technology and practice of AI incident cataloging. These contributions are motivated to both explain our practices and to inform others that may be looking to expand the network of actors collecting and sharing AI incidents.

\section{The AI Incident Database and Related Work} \label{related_work}

Incident reporting is a practice common in many other fields, from aviation \cite{faa} to environmental monitoring \cite{epa}.
Perhaps the field with established incident reporting practices most relevant to AI is cybersecurity, as both fields are rooted in monitoring wide-spread digital technologies where software is associated with both tangible harms (e.g. compromised autonomous vehicle systems resulting in accidents) as well as intangible harms (e.g. compromised data storage resulting in disclosure of private information). Importantly, there exists a widely-adopted practice to handle potential incidents shared by research, industry, and government actors alike, realized in the Common Vulnerabilities and Exposures (CVE) Program \cite{mitre_cve}. The CVE Program catalogs reported cybersecurity vulnerabilities for verification and awareness across many stakeholders and industries, providing a ``canonical'' identification of a vulnerability so that it may be consistently referenced.

\paragraph{The AI Incident Database.} Inspired by the CVE program and other incident reporting practices, the AI Incident Database (AIID) catalogs AI incidents with the goal of reducing the likelihood that similar incidents recur \cite{aiid_paper}. The AIID hosts a human-curated dataset of AI incidents, each of which are substantiated by one or more reports. Currently there are over 750 distinct AI incidents detailed by more than 3000 indexed third-party reports.
The nature of the incidents spans various AI systems and contexts, ranging from harms associated with autonomous vehicle accidents to algorithmic discrimination. The AIID's incidents are maintained by editors who have spent years developing guidelines\footnote{\url{https://incidentdatabase.ai/editors-guide/}} in response to changing technologies, novel harms, and frequent discussion. The resulting dataset, which is under continual development, is publicly available to download as a database snapshot.\footnote{\url{https://incidentdatabase.ai/research/snapshots/}}

In addition to serving as a public educational resource on AI safety and ethics \cite{feffer_ai_2023}, the AIID supports deeper AI incident research by providing a platform for further annotating incidents according to different domain-specific analyses and research goals. These are broadly referred to as \textit{taxonomies} in the AIID platform. The two most prominent taxonomies hosted on the AIID are the Center for Security and Emerging Technology (CSET) AI Harm Taxonomy and the Goals, Methods, and Failures Taxonomy, which each address distinct incident analysis goals.

The CSET AI Harm Taxonomy for the AIID \cite{cset_taxonomy} applies a conceptual framework that \emph{operationalizes notions of harm}. The framework and taxonomy identify differences between ``tangible'' and ``intangible'' harms, provide categories of these harms (e.g. harm to physical health or safety, human rights violations, financial loss), and distinguish harms that have occurred versus those that may occur in the future, among other features.
At the time of writing, 214 incidents in the AIID contain classifications in the CSET AI Harm Taxonomy for the AIID, but the framework is intended to be illustrative of what is possible in incident annotation more broadly than solve annotation problems within the AIID.\footnote{More details about the taxonomy’s annotation methodology and application to AIID can be found in a publicly-available guidelines: \url{https://github.com/georgetown-cset/CSET-AIID-harm-taxonomy}}.

The Goals, Methods, and Failures (GMF) taxonomy \cite{gmf_paper} classifies the same underlying incidents in the AIID, but it focuses on failure cause analysis by interrelating the system's goals, its methods, and their technical causal factors for the observed failure events. The taxonomy structure and methodology encourage \emph{considering what is known or observed} of these aspects along with what is \emph{potential or likely}, guiding how to apply and interpret expert technical knowledge about AI failures in the presence of uncertainty. These ``confidence modifiers'' are relevant when the incidents in question are based on public but incomplete third-party reports. At the time of writing, 188 incidents in the AIID contain classifications in the GMF taxonomy.

The efforts behind cataloging incidents and applying these taxonomies present operational challenges beyond the scope of producing research papers. These are open-ended projects for which we expect to never arrive at an end state. As such, we share the operational concerns that companies and countries may also have, herein presented as our research question \textbf{(RQ)}: what are the current operational challenges to reporting, indexing, and analyzing AI incidents learned from the AI Incident Database as well as the application of its long-term supported taxonomies? To explore the question, we identify incidents in the underlying AIID dataset that ``stress'' its editing methodology and propose qualitative themes that characterize these cases. We then explore how the downstream taxonomies that further annotate incidents in the AIID either handle or are challenged by these. Throughout, we discuss the implications of these findings as lessons for future AI incident reporting practices.

No such widely-shared and successful practices exist for AI incident reporting despite the growing legislative acts and drafts that would direct agencies to develop AI incident reporting programs.
The term \textit{``AI incident''} itself is a relatively new term that presents definitional challenges and ambiguities. Beyond inheriting the issue of defining what is considered to be ``AI'' across both policy and technical domains \cite{krafft_defining_ai_2020}, defining what can be considered an AI \textit{incident} touches fundamental questions of semantics, values, and law.

Multiple policy-oriented definitions published in 2024 require that an AI incident identifies harm to people, rights, property, or the environment. For example, the definition of ``serious incident'' in the EU AI Act \cite{eu_ai_act} or definition of  ``AI incident'' by the Organization for Economic Co-operation and Development \cite{oecd_ai_incident} both position ``incidents'' in relation to their harms. Other perspectives argue a broader sociotechnical approach where AI incidents should be identified as ``algorithmic troubles,'' events caused by particular algorithmic systems that provoke anxiety or concern about the nature of their functioning \cite{meunier_troubles}. Prior works have focused on similar concerns under the term ``AI accidents'' \cite{cset_ai_accidents}, though this term is less popular and nominally excludes purposeful harms.

The construction of these definitions opens up questions integral to reporting practices, such as how to define or scope the nature of harm. An OECD report on approaches to defining AI incidents identifies various potential dimensions of harm – such as the type of harm, severity, reversibility – and highlights difficulties in aligning with existing technical standards and terminology related to severity and risk \cite{oecd_stocktaking_2023}. \citet{cset_taxonomy} argue that assessing intangible or derivative harms of AI incidents is inherently difficult, and the difficulty of identifying harm across different incident reporting efforts can lead to inconsistent measurements and recommendations.  

\subsection{Actual harm vs. potential harm}
The AIID maintains an operating definition of an AI incident in its editor guidelines as \emph{``an alleged harm or near harm event to people, property, or the environment where an AI system is implicated.''}
Notably, this articulation includes ``near harm'' events as AI incidents, which characterize incidents such as Incident 8: \textit{``Uber Autonomous Cars Running Red Lights''} \cite{aiid:8}.
This construction captures behaviors of AI systems that would have caused harm if not for other environmental factors – a particular subset of alleged ``potential'' harm.
Complementary to this term and intended to capture other types of potential harm, the AIID also ingests what it terms as ``AI issues'', which are \textit{``an alleged harm or near harm by an AI system that has yet to occur or be detected.''}
In practice, AIID’s AI issues are often ``incidents waiting to happen'' and are collected evidence that understanding of where AI presents the potential for harm.
This articulation of \textit{actual} versus \textit{potential} harm is also operationalized in the OECD’s AI incident definition and guidelines, which propose the term ``AI incident hazard'' to capture analogous types of potential and future harm caused by AI systems. The term ``hazard'' is more common in the product safety community and is likely to be the dominant term moving forward. In the future, we will be harmonizing the AIID ``issues'' with the OECD ``hazards.'' 

\subsection{Reporting and Monitoring Projects}

In addition to the AIID, other projects that seek to index AI incidents include the OECD AI Incidents Monitor (AIM)\footnote{https://oecd.ai/en/incidents} and the AI, Algorithmic, and Automation Incidents and Controversies (AIAAIC) repository,\footnote{https://www.aiaaic.org/} both of which index similar public reports of incidents albeit by different methodologies or tools. More recently, incident databasing activities have begun to target specific types of harms and incidents, such as the Political Deepfakes Incidents Database \cite{pdid_2024}, or specific domains, such as the Database of AI Litigation (DAIL)\footnote{https://blogs.gwu.edu/law-eti/ai-litigation-database/}. Other related efforts focus on aggregating and recording AI system vulnerabilities, such as the AI Vulnerability Database,\footnote{https://avidml.org/} also inspired by the CVE Program.

\subsection{Taxonomies for Deeper Analysis}

Detailed analysis and organization of AI incident data and AI risks more broadly is often done using taxonomies or other structured annotation. These taxonomies vary in their purposes, often targeting specific technical AI research and development, public policy, humanities-oriented, or educational audiences. 

AI incident- and harm-specific taxonomies can seek to generally annotate incidents and harms and opt for inclusion, such as the aforementioned CSET AI Harm and GMF taxonomies. Other taxonomies focus on risks or harms implicating particular use cases of systems, such as speech generation system harms \cite{hutiri_2024}, or analyze specific contexts in how the systems are used, such as ``misuses'' of generative AI \cite{marchal_misuse_2024}. Other recent works have compared AI incident- and harm-focused taxonomies in more detail and propose additional taxonomies that seek to annotate AI incidents for broader public consumption \cite{abercrombie_2024}.

The abundance of taxonomies seeking to characterize risks, harms, and incidents caused by AI and other algorithmic systems is so great that meta-studies of the recent literature seek to compose more abstract or more systematically inclusive taxonomies. Based on a scoping review of 172 relevant papers and resources, Shelby et al. proposed a high-level taxonomy of sociotechnical harms and reflect on challenges in unifying taxonomy development and applications across various computing disciplines \cite{shelby_taxonomy_2023}.

Nevertheless, despite the awareness raised by incident cataloging projects, the challenge of capturing the dynamic nature of the AI incidents remains. Further taxonomic analysis and techniques applied to incident data may be required to better understand and prevent future harms \cite{turri_why_2023}. In the following section, we present the challenges of ingesting and organizing incident data, especially as it impacts the ability to aggregate and analyze data.

\section{Editorial Challenges to Indexing AI Incidents}

We tentatively group challenging editing themes as ``key AI incident editing challenges.'' These themes are not mutually exclusive; AI incidents that challenge intuitive definitions often exhibit one or more of these patterns in practice.

Finally, we review how these challenges impact downstream analysis of AI incidents, focusing on the CSET AI Harm Taxonomy and the GMF Taxonomy.

\subsection{Key AI Incident Editing Challenges}

\subsubsection{(1) Temporal Ambiguities.}

While some harms caused by AI systems are the results of discrete events with defined and intuitive event timelines, other AI incidents can span ambiguous amounts of time, present as series of events that are difficult to neatly separate and aggregate in a database setting, or obscure when exactly harm can be considered to have occurred. This challenges incident reporting models that require knowing the exact or even approximate times of harm occurrence, which are desirable when seeking to analyze incident trends. 

A well-known example of an AI incident with temporal ambiguities is known as the ``Dutch Childcare Benefits Scandal,''
wherein Dutch families were wrongfully accused of tax fraud in part due to a discriminatory algorithm \cite{aiid:101, Peeters_dutch_benefit_scandal}.  In this incident, it is unclear precisely when the problematic algorithmic components in the decision-making process were introduced, which would constitute a potential ``start'' of the incident. It is also difficult to ascertain when the harms of the incident can be considered to have ``ended'' given that the harm to families was deep and social; as a result of the algorithmic fraud accusations, many children were taken into foster care and the financial burdens were severe. How does an AI reporting system or database capture the complexity of such a timeline of harms? How can ambiguous records be used in data-driven approaches to regulating AI?
There is a tension between unifying all of the harms experienced under a single umbrella event and considering the harm to each family or victim as its own distinct AI incident.

It is also clear that AI incidents can be \textit{ongoing}. Those that are a part of disinformation campaigns that leverage AI technologies across platforms, for example. In these situations, the individual pieces of disinformation often defy simple categorization as they evolve over time and reach across multiple jurisdictions while often being described as part of wider, sustained influence operations.

For instance, incidents like
\textit{``Kremlin-Linked Entities Allegedly Using Generative AI to Spread Russian Disinformation in Latin America''} \cite{aiid:585},
\textit{``Russia Using AI in Disinformation Campaigns to Erode Western Support for Ukraine''} \cite{aiid:602},
and \textit{``TikTok AI System Used to Amplify Election Disinformation by Foreign Networks''} \cite{aiid:761}
are not isolated events; they certainly contain many discrete harm events, but are often discussed collectively as long-running problems that are continuously adapting and expanding.

Reporting on incidents of this nature involves navigating the inherent tension of capturing the specificity of each incident while also acknowledging the overarching nature of these disinformation campaigns. In some cases, sufficient reporting exists on a singular harm event to formulate one incident ID, such as \textit{``Deepfake Targets Olena Zelenska in Russian Disinformation Campaign''} \cite{aiid:755},
but in the aforementioned cases, reporting the wider contours of ongoing and long-running incidents becomes necessary. That complexity is further compounded by the ways in which editorial decisions at the journalistic level might merge incidents into larger narratives in such a manner that eschews the detailed specifics that are necessary for discrete incident reports.

\subsubsection{(2) Multiplicity.}

Most AI incidents are not unique and unreproducible events but are often caused by systems that are involved in our everyday lives. This means that patterns of harm events can develop wherein a fixed AI system can harm different parties in nearly the same causative manners or repeat the same failures. For example, there are multiple distinct entries for autonomous vehicle accidents implicating the same vehicle models and systems. While the causative factors can differ between incidents (e.g. unexpected highway braking \cite{aiid:434} vs. stop sign misidentification \cite{aiid:145}),
it is often the case that each incident report bears no distinguishing technical details, resulting in reports of harm and investigations that group together such accidents without a means to distinguish individual harm events \cite{aiid:711}.

Other patterns of AI incidents can be more trivially repeatable. Consider the potential for widely-available machine translation systems to separately fail many individuals in the same or similar manners, such as perpetuating gender bias \cite{aiid:59} or creating offensive translations \cite{aiid:415}.
The same can be said for popular large language model (LLM) applications which have millions of monthly active users, for which even a relatively small probability of model ``hallucination'' can materialize to impact thousands of individuals.

In this way, \textbf{AI incident reporting systems require a strategy to handle multiplicity of incidents,} especially if the system has a mandate to record every instance of harm (as opposed to being selective). As a potential way to index such incidents, the AIID proposed identifying and cataloging ``AI incident variants'', defined as an \emph{``AI incident that shares the same causative factors, produces similar harms, and involves the same intelligent systems as a known AI incident''} \cite{aiid_variants}. The need for incident reporting strategies to relate or cluster similar incidents and harms in a lightweight and performant manner is increasingly important, especially as popular and generalizable AI systems are increasingly made available and deployed as commercial products (such as LLMs).

Aside from generalized AI incident reporting systems, the multiplicity of certain types of AI incidents also \textbf{motivates the opportunity for more specialized incident reporting and database practices that restrict and apply domain expertise to particular types of incidents.} For example, the Political Deepfakes Incidents Database (PDID) focuses explicitly on synthetic media that relate to political and public discourse given the proliferation of such deepfakes and the generative AI systems that produce them \cite{pdid_2024}.

\subsubsection{(3) Aggregate and Societal Harms.}

All of the AI incident definitions discussed would generally consider harms to groups of people caused by AI systems as incident-worthy. But what are AI incident reporting systems to do when the harm in question is an emergent effect distributed across thousands of people and potentially unobservable at the individual level? Put differently, what happens when an AI system is seen to have harmed society? Such incidents are perhaps among the most pressing to rigorously identify but also the most difficult, as it is often the case that ``aggregate'' and society-impacting incidents do not indicate tangible harms to individuals and require a holistic perspective. In such cases, the AIID opts to take this broader perspective and index such incidents, but such incidents are perhaps the most difficult to analyze and compare with others.

Algorithmic recommendation systems are particularly of interest, as they are often deployed globally and are potential sources of societal harm such as reinforcing existing inequalities \cite{saurwein_automated_2021} or amplifying misinformation and extremist content \cite{whittaker_recommender_2021}. Consider Incident 348: \textit{``YouTube Recommendation Reportedly Pushed Election Fraud Content to Skeptics Disproportionately,''} where YouTube's recommendation algorithm allegedly disproportionately displayed election fraud content to users most skeptical of the election's legitimacy compared to less skeptical users \cite{aiid:348}. While it is unclear if individuals receiving these recommendations can be considered directly harmed, there is a considerable danger to society presented by the associated system if we maintain that polarizing political discourse online is a harm to society in itself.

Despite the ambiguity present in such an incident, other society-impacting incidents can positively indicate individual harms. Consider again the Dutch Childcare Benefits Scandal \cite{aiid:101}, wherein specific individuals and families were directly impacted by algorithmic decision-making, and the scale of which was so significant that it resulted in resignation of the then executive government \cite{Peeters_dutch_benefit_scandal}.

As incident reporting practices formalize, \textbf{they must consider how they will substantiate and investigate harm to groups of people and society absent clear harm to individuals.}

\subsubsection{(4) Epistemic Uncertainty in Incident Reporting.}

In the prior section, particular dimensions of ambiguity can be considered to be artifacts or flaws introduced by the way AI incidents are defined, modeled, and compared within incident reporting systems. However, the role of uncertainty in incident reporting is more fundamental than artifacts introduced by modeling an incident for the sake of indexing it. Rather, uncertainty about facts of AI incidents can be \emph{epistemic} – the knowledge of whether certain facts about an AI incident are true or false is unavailable or unattainable. Such facts might concern the nature of the implicated AI system (such as exact machine learning model weights or configuration of the broader AI system), downstream impacts (such as the number of people definitively harmed or exposed), or any number of other details.

Much of this information is simply \emph{unavailable} from the perspective of the incident reporter or collector. As the AIID relies primarily on ``open source'' incident information to identify incidents (i.e. that which is publicly available, such as third-party media reports, legal allegations, and other public disclosures), editors frequently encounter reports of incidents that lack insight into particular facts \cite{gmf_paper}.

Further, while it is generally true that developers and deployers of implicated AI systems naturally have more privileged access and knowledge about technical details and operating contexts, their knowledge of relevant facts is not necessarily perfect or even sufficient. Key technical information (such as AI system architecture or model weights) may not be fixed in particular AI systems that allow downstream users to customize aspects of the system or integrate it with other systems (e.g., for large language models that operate on a context window of data provided by the users); operational information (such as device logs of system inputs/outputs) may not sufficiently monitor the system or be maliciously destroyed or corrupted. What are AI incident reporting systems to do when information is – for all intents and purpose – irretrievably lost? In this way, AI incident information can be \textit{unattainable}. This illustrates that even under a mandatory reporting regime, an AI incident reporting system will grapple with missing information to some extent.

Another example of re-occurring (and perhaps unavoidable) uncertainty in AI incident analysis is establishing \textit{causality} in the link between the AI system and the harm event. That is, the harm event in question would not have occurred if not for the behavior of the AI system implicated. This construction of causality is inherently counterfactual and here the answer can also be considered \textit{unattainable} and subject to epistemic uncertainty.

Regardless, such uncertainty should not deter AI incident reporting practices – just as similar uncertainty has not stopped aviation or cybersecurity incident reporting. \textbf{Rather, the takeaway is that AI incident reporting systems must embrace methodologies and cultures that are conducive to investigation in the presence of uncertainty.} While the core AIID editing methodology does not explicitly address such uncertainty, downstream taxonomies and consumers of this information can apply additional methods that do so (e.g., ``confidence modifiers'' in the GMF Taxonomy, discussed next).

\subsection{Impacts on and Further Lessons From AI Incident Taxonomies}

Not only do the aforementioned challenges impact the decision of whether and how to include candidate AI incidents in the database, they also impact the ability of third-party taxonomies in the AIID platform to analyze the data. Depending on the particular methods and goals of a taxonomy, these issues can be mitigated or they can compound.

\subsubsection{CSET AI Harm Taxonomy.}

A goal of the CSET AI Harm Taxonomy for AIID\footnote{\url{https://incidentdatabase.ai/taxonomy/csetv1/}} is to create a structure for extracting AI harm information that helps AI stakeholders to understand risk and harm from AI systems, track trends in AI harm incidents, and identify emerging types and vectors of AI harm \cite{cset_taxonomy}. The taxonomy differentiates between tangible and intangible harm. The taxonomy requires tangible harm to have observable injury, loss, or damage. In contrast, intangible harm usually cannot be directly observed and often depends on societal norms or cultural context. While the taxonomy attempts to capture details about the AI system, sector, environment, entities, locations, dates, and type of harm that were involved in the AI incident, it was primarily developed to identify if tangible or intangible AI harm was involved; distinguish between harm events, near-misses, and issues; and decrease inter-annotator variation.

In addition to confirming the key modeling and uncertainty issues in AI incidents, key lessons learned from the development and application of the CSET AI Harm Taxonomy include acknowledging the challenge of mitigating \textbf{inter-annotator variation}. Inter-annotator variation can be reduced but not removed through iterative development of definitions, examples, and annotation processes. For instance, due to the iterative development process, the CSET AI Harm Taxonomy includes questions about the domain or conditions in which the potential AI harm occurred and sought to focus annotators on essential details. These questions helped reduce inter-annotator variation but did not eliminate it, in part because each annotator brings their distinct cultural and personal experiences.

In addition, the taxonomy's focus on differentiating tangible and intangible harms highlights a particular source of uncertainty. Especially in the case of \textbf{recording intangible harms}, it is inherently difficult to know if intangible harm definitively occurred or if there was simply a nonzero probability that it could have occurred.

\subsubsection{Goals, Methods, and Failures (GMF) Taxonomy.}

As a failure cause analysis taxonomy, the GMF taxonomy\footnote{\url{https://incidentdatabase.ai/taxonomy/gmf/}} seeks to interrelate the goals of the system deployment, the system’s methods, and technical causal factors for the observed failures in AI incidents. Notably, the GMF taxonomy provides multiple features that intend to mitigate the effects of uncertainty in AI incident reporting. Firstly, the taxonomy records both rich snippets of text from reports as well as annotator rationale for resulting annotations, which serve to address potential uncertainty introduced by public but incomplete third-party reports.

Secondly, the taxonomy also specifically identifies which annotations represent ``potential'' or ``known'' facts of the incident. In this way, the GMF taxonomy \textbf{explicitly addresses epistemic uncertainty.} Downstream consumers of GMF annotations can compare and aggregate incidents with respect to these confidence modifiers, and these modifiers enable a process where expert and editor communities can iteratively review and upgrade (or downgrade) annotations \cite{gmf_paper}.\footnote{Today, the GMF taxonomy powers the AIID's ``Risk Checklists'' feature, which enables users to retrieve incidents by potential goals, methods, and failures of underlying AI systems in order to illustrate risks.}

While the taxonomy mitigates uncertainty by explicitly recording it, it still faces certain ambiguities. In GMF, temporal ambiguity issues transfer from incident delineation to the failure cause analysis domain. Incidents that span an extended period of time may involve multiple AI systems that change over time (e.g. different model iterations, updates to problem-solving approaches, or new mitigation measures), which may in turn result in new methods to annotate or sources of technical failure to consider (it is less likely that the relevant goal of the AI system will change). As a result, incidents that exhibit temporal ambiguity increase complexity of failure cause analysis.

In addition, the development and application of the GMF taxonomy to the AIID dataset surfaces the concern of which incidents are ``in scope'' for its analysis. The GMF taxonomy focuses on technical failure causes (i.e. methodological or implementation-related steps) that lead to the manifestation of harm. This is in tension with fact that the AIID also indexes incidents where the application of AI systems leads to harm without technical failings. This not only includes cases of malicious use of AI systems (e.g. ``deepfakes'' \cite{aiid:544}) but also extends to incidents that have aggregate or society-level harms that are not attributed to technical failures (as discussed previously). A simple takeaway here is that the GMF taxonomy and others need not be applied to such incidents, but this highlights the variability of incident editing methodologies where the underlying dataset is more broad.\footnote{Ongoing discussion of how to discern non-technical failure causes in the GMF Taxonomy can be found here: \url{https://github.com/responsible-ai-collaborative/aiid/issues/2704}}

\section{Conclusion}

This study presents operational insights into the challenges of cataloging AI incidents faced by the AI Incident Database. They are rooted in seven years of deliberative work to faithfully describe AI incidents for researchers, developers, policy makers, and the general public.
We highlight potential mitigations and solutions to these challenges with the goal of empowering the growing AI incident community.
Just as the development of AI is dynamic, so must be AI safety practices. AI incident reporting will continue to confront new ambiguities with new technologies, but safer systems will become possible by a careful examination and presentation of the past.

\section{Acknowledgments}

This work benefits from years of discussion and concept development among a collection of incident and taxonomy editors. We are particularly indebted to Mia Hoffman, Patrick Hall, Khoa Lam, Zachary Arnold, Helen Toner, Janet Schwartz, and Kate Perkins, but we also would have much less data with which to work if not for the 96 (and counting) named submitters to the AI Incident database submitting the reports of 3,300+ on AI incidents across 1,200 different source domains. Finally, we are lucky to work with a talented team of engineers building the computer systems behind the AIID, including César Varela, Pablo Costa, Clara Youdale, and Luna McNulty.

\bibliography{manuscript}

\end{document}